\documentclass[twocolumn,prb,showpacs,preprintnumbers,amsmath,amssymb,a4paper,floatfix]{revtex4}

\usepackage{graphicx}
\usepackage{dcolumn}
\usepackage{bm}

\begin{document}

\preprint{APS/123-QED}

\title{Pt steps decorated by 3d nanowires calculated via an order-$N$ method}

\author{Sam Shallcross}
 \altaffiliation{Physics Department,  Link\"oping University.}
 \email{phsss@ifm.liu.se}

\author{Igor Abrikosov}
\affiliation{Physics Department, Link\"oping University.}

\author{Velimir Meded}
\affiliation{Physics Department,  Uppsala University.}

\date{\today}

\begin{abstract}

We present the magnetization energy and magnetic moments of 3d monowires aligned
along the step edges of Pt (533) and (322) vicinal substrates. We employ an electronic
structure method which scales linearly with the size of the system. This allows us to treat
with ease both ferromagnetic and antiferromagnetic solutions, and in principle the
method should allow for the evaluation of more complex systems, such as wires deposited 
on rough step edges. We find that Co, Fe, and Mn are almost perfect Heisenberg systems, with
almost no change in the moment between antiferromagnetic and ferromagnetic solutions. A
large polarisation of the Pt substrate is also observed. Our results are in close agreement with recent
experimental results obtained for the case of a Co monowire.
As expected the trend is for the late d series to
be stabalised by ferromagnetic long range order and the middle of the d series by
antiferromagnetic order. Since our method is only \emph{locally} self consistent we
discuss as some length the convergence of the results with respect to the size of the
region which is treated self-consistently.
\end{abstract}

\pacs{Valid PACS appear here}
                             

\maketitle

\section{Introduction}

Physics on the nanoscale has developed enormously in the last decade and
the construction of a wide variety of nanoscale systems has become possible. This
usually proceeds through either an atom by atom manipulation using STM, or by
growth on a substrate favorable to the formation of the desired nanostructure.
Stand out amongst the latter category of experiments has been the production
of parallel monoatomic chains aligned along the step edges of a Pt vicinal substrate
\cite{Gambardella02}. Vicinal substrates, created by the controlled miscut of the crystal 
off a high symmetry plane, present an ordered array of step edges and thus form
an ideal template for the growth of monoatomic wires.

One of the fundamental issues which arises in the context of nanophysics is how
very well known phenomena such as magnetism are altered in systems of
reduced dimensionality. A variation in the magnetic properties of the 3d and
4d transition metals as a function of dimensionality and substrate type has
been observed on several occasions \cite{Blugel92a,Blugel92b,Wu92,Lang94,Stepanyuk96,
Bellini01,Spisak02,Lazarovits03},
where the systems investigated have ranged from zero-dimensional adatoms, monowires,
and monolayers. The picture that has emerged from these investigations is that
elements which in the bulk situation are non-magnetic may develop large magnetic
moments in the nanoscale. More precisely it has been observed that
the peak of the moment-valence plot shifts to lower valence as the dimensionality 
is reduced.
However, the early prediction of the magnetism of two dimensional
monolayers of 4d metals was not realized in many cases. This was due to the
sensitivity of the magnetism to structural defects and relaxation, these tending
to reduce the magnetic moment from that of the ideal structure.
On the other hand the recent investigation of the magnetic nature
of the 3d metal Co deposited in
one-dimensional chains on the Pt$(997)$ surface \cite{Gambardella02} showed
a dramatically enhanced spin and orbital moment.

From the theoretical point of view one of the problems with the study of 
low dimensional systems is that the reduction in symmetry and the
corresponding increase in system size mean that the calculations can become
quite demanding. This is simply due to that fact that most of the theoretical
approaches scale as $N^3$, with N the number of inequivalent sites. Alternative
{\it ab initio} techniques with more favorable scaling do exist however. In particular,
Bellini {\it et al.} \cite{Bellini01} have determined the magnetic
structure of 4d monoatomic rows on Ag vicinal surfaces, using an {\it ab inito}
method that scaled as order $N$ with the number of substrate layers. However
the scaling with the number of inequivalent wire atoms remained $N^3$. This, as
the authors noted, meant that calculations involving inequivalent wire atoms 
(needed for the calculation of antiferromagnetic wire configurations) 
became again very demanding. On the other hand, Eisenbach {\it et al.} have
recently deployed a genuine {\it ab inito} order $N$ method in the study of 
Fe monowires embedded in a Cu matrix. 
Using this approach they were able to determine the magnetic structure of
the Fe wire.

{\it Ab inito} order $N$ methods clearly hold a lot of promise for the investigation
of systems on the nanoscale. With the ability to calculate systems of several hundred
atoms one can consider, for example, magnetic clusters either embedded within the bulk
of another material or deposited on a substrate. However, the favorable scaling
of order $N$ methods comes at a certain price of accuracy. This is due, in the case of the
method deployed by Eisenbach {\it et al.} and to the related method used by us in this work,
to their being only \emph{locally self consistent}. Hence, it is necessary to establish
that the magnitudes of quantities one is interested in are within the scope of the
method.

In this work we investigate the magnetic properties of 3d nanowires deposited on 
the step edges of a Pt vicinal surface. We consider both ferromagnetic (FM) as
well as anti-ferromagnetic (AFM) magnetic configurations, and establish the
accuracy of the method for the relevant magnetization energies. Although the
calculations we perform here do not have spin-orbit (SO) coupling this will not be 
important for the determination of energy differences between FM and AFM magnetic configuations,
which will be order the order of mRy and hence orders of magnitude bigger than the magnetic
anisotropy from the SO interaction.

\section{Vicincal geometry and supercell}

A vicinal surface is one created by miscutting away from a high symmetry plane. These
surfaces neccesarily display a periodic array of steps aligned along some specific direction, separated
by terraces of the high symmetry plane the surface was miscut from.
We will consider here a surface miscut off the high symmetry ($111$) plane. In this
case there are two possible vicinal surfaces with steps that have either ($100$) 
or ($\bar 111$) microfacets. These two families of surfaces are described by the
Miller indices ($p+1,p-1,p-1$) and ($p-2,p,p$) respectively with $p$ being the number
of atomic rows in the terraces separating steps.

The template that we use for depositing the 3d monowires on will in this study be
the ($p+1,p-1,p-1$) vicinals, and we consider surfaces with 2 atom wide and 4 atom wide
terraces. As usual we replac the semi-infinite geometry
with that of a supercell which is then periodically repeated in space. This leaves us with
a slab with two surfaces, one of which we will decorate with a monowire. It is important
that these slabs should not interact and also that the two surfaces should not interact
across the slab.
Here we accomplish this by separating the slabs which are 8 ($111$) layers in thickness by 
7 ($111$) vacuum layers. This leads to supercells of 25 atoms and 55 atoms for the
($311$) and ($533$) vicinals respectively. In order to consider the possibility of
anti-ferromagnetic solutions of the monowire we simply double the supercell in the wire
direction, leading to supercells of 50 and 110 atoms.

\section{The order-$N$ locally self consistent Green's function method}

The Locally Self-consistent Green's Function method (LSGF) \cite{abrikos97} 
is based on an approach whereby a supercell of $N$ sites is divided into $N$ separate 
\emph{local interaction zones} (LIZ) each of $M$ sites. The motive for this construction
is the observation that that the properties of the central site in each LIZ will approach
those that would be found by a full solution of the supercell as $M \rightarrow \infty$.
This is very similar to the Locally Self-consistent Multiple Scattering (LSMS) method 
used by Eisenbach {\it et. al} in their work on Fe wires embedded in Cu \cite{eis02}. 
Essentially, the only difference between the two approaches is that in the LSMS method
each LIZ is embedded within a free electron gas whilst in the LSGF method one embedded
the LIZ into some judiciously chosen effective medium.

Formally this is achieved by writing a Dyson equation for the central site of each LIZ

\begin{equation}
g_{{\bf R}{\bf R}} = \tilde g_{{\bf R}{\bf R}} + \sum_{{\bf R'} = 1}^{M} 
\tilde g_{{\bf R}{\bf R'}} (\tilde P_{\bf R'} - P_{\bf R'}) g_{{\bf R}{\bf R}}
\end{equation}

By solving this equation for every site in the supercell one finds the diagonal element of
Green's function matrix $g_{{\bf R}{\bf R}}$ for all sites, and from this set
one can construct the effective medium Green's function $\tilde g_{{\bf R}{\bf R'}}$. This
process is then iterated to self-consistency being first initiated by some suitable guess.
One must note that although the full Green's function matrix $g_{{\bf R}{\bf R'}}$ will
not correspond to that of the supercell under consideration, provided $M$ is large
enough, the site diagonal element will.
Since it is the only the site diagonal
elements that are required to determine the local density of states for each site
one may extract reliable information on charge density, one electron energy, and total
energy of the supercell from this method.

The effective medium plays only the role of a boundary condition for the LIZ, but convergence
with LIZ size will be strongly affected by this choice \cite{abrikos97}. In the case of
a highly inhomogeneous structure one way to proceed is by making the effective medium
more complicated (for example by allowing for inequivalent effective atoms), however in
the case of a vicinal surface this is not practical. (Since the number of inequivalent vicinal
layers is equal to the number of sites in the supercell.) We therefore proceed by the use of
the simplest single site effective medium construction. The criteria that
such an effective medium should sattisfy in such a case is that the "effective atom"
most resembles the maximum number of sites. With only one type of effective medium site this
is clearly not possible in the case of seperate vacuum and metal regions in the supercell.
A way that suggests itself is to sattisfy the two natural limits by having an effective
vacuum atom and an effective metal atom. Near the interface region the size LIZ will need
to be large to eliminate the error at the central site, but away from the interface region
the effective medium will come to resemble more and more closely the real material and hence
the error with the same size of the LIZ is smaller.

For simplicity we choose here a single site effective medium constructed from all metal sites via the
average $t$-matrix approach. The rational behind this can be seen by considering the LIZ 
around an atom in the surface. The error committed for vacuum sites in the LIZ should be
less important, as the density of states there is anyway small, while the error for
the metal sites will be less for not having a spurious vacuum term in the effective medium,
and indeed should vanish in those sites which no longer feel the perturbation of the surface.

One must emphasize that there is no \emph{direct scattering} between distant sites in
this method. Beyond the LIZ boundary the effective medium is single site and contains
only information related to the underlying lattice and no information whatever about 
the geometry of the vicinal surface. Hence quantum effects due entirely to such
direct scattering processes with be lost. This may be of importance when considering,
for example, step-step interactions.

\section{Computational details}

The method deployed here was a real space technique based on the Korringa-Kohn-Rostoker 
Green's function scheme in the atomic sphere approximation. The basis consisted of $s$, $p$, and
$d$ orbitals, and corrections to the spherical charge density were included up to $l_{max} = 4$.
The calculations were performed in the Local Density Approximation (LDA) with the Perdew, Burke,
and Ernzerhof parameterization of the results of Ceperly and Alder. 
The integration of the Green's function
was taken in the complex plane with 16 energy points on a semi-circular contour.
The lattice parameter for Pt was taken to be the theoretical LDA value.

\section{Convergence of total energy and magnetic moment with LIZ}

\begin{figure}[floatfix]
\caption{Magnetic moments of 3d monowires and total moment of substrate and vacuum sites for
monowires on Pt $(311)$ (full symbols) and  Pt $(533)$ (open symbols). Results shown are for $LIZ = 5$.}
\vspace{7mm}
\begin{center}
\includegraphics[angle=-00,width=0.45\textwidth]{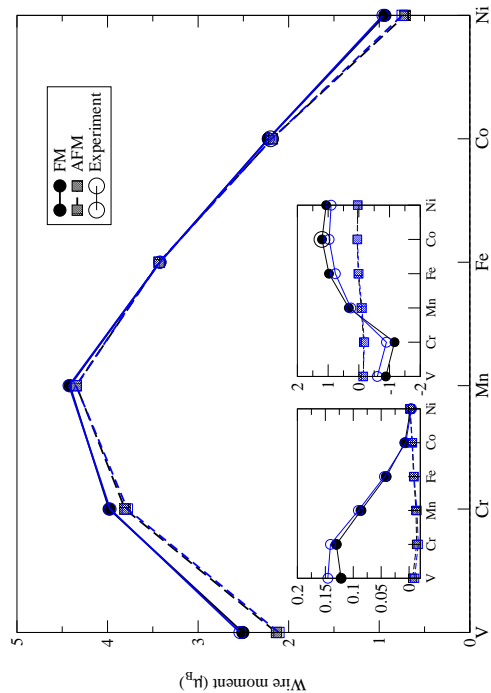}
\end{center}
\label{fig:moments}
\end{figure}

\begin{figure}[floatfix]
\caption{Average magnetic moment per principle layer for 3d monowires
on Pt $(311)$ (full symbols) and  Pt $(533)$ (open symbols). Results shown are for $LIZ = 5$.}
\vspace{7mm}
\begin{center}
\includegraphics[angle=-00,width=0.45\textwidth]{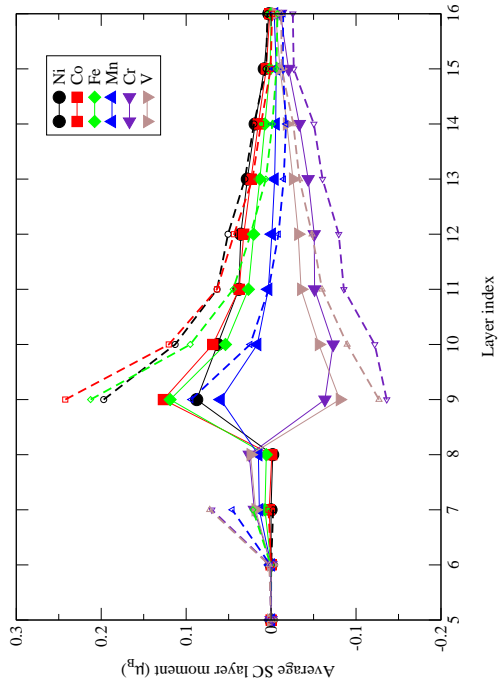}
\end{center}
\label{fig:momRS}
\end{figure}

The success of the LSGF method as a genuine order-$N$ method for arbitrary structures
depends of course on how quickly is the convergence with respect to the LIZ size. By construction
the rate of convergence of the supercell will depend solely on the convergence of the individual
LIZ's. This in turn depends on how well the effective medium corresponds to the system at hand.
A consequence of this is that the convergence of the total energy should become 
worse linearly with increasing $N$. In Table \ref{tab:convTE} are shown the total
energies for the supercell consisting of a Co monowire deposited on Pt ($533$). From
this one may conclude that a LIZ of 55 atoms is enough to capture the energetics
of this system. In Table \ref{tab:convmE} are shown the wire and substrate magnetic
moments and magnetization energies. The convergence is better for the wire than
for the substrate quantities. For the moments this can be understood as due to
the fact the the 3d monowire will be a very good local moment system whereas the Pt
substrate is of more itinerant character. The more localised the magnetic moment the better 
it is expected to converge in real space.

\begin{table}
\caption{\label{tab:convTE} Total energies for FM and AFM Co wires on Pt($533$) substrate. Also
shown is the difference $\Delta E = 2 E_{FM} - E_{AFM}$}
\begin{ruledtabular}
\begin{tabular}{lccc}
LIZ (sites)  & E$_{FM}$ (Ry) & E$_{AFM}$ (Ry) & $\Delta E$ (Ry) \\ \hline
 13   &  -19492.853536 & -19492.853536 & -0.045265 \\
 19   &  -19492.853572 & -19492.853572 & -0.014410 \\
 43   &  -19492.853573 & -19492.853573 & -0.006985 \\
 55   &  -19492.852730 & -19492.852730 & -0.009845 \\
 79   &  -19492.852714 & -19492.852714 & -0.010615 \\
 87   &  -19492.852758 & -19492.852758 & -0.010120 \\
\end{tabular}
\end{ruledtabular}
\end{table}

\begin{table}
\caption{\label{tab:convmE} Wire and substrate moments for FM Co wire on Pt($533$) substrate. Also
shown is the site projected energy for the wire site, and the sum of
site projected energies for the substrate sites.}
\begin{ruledtabular}
\begin{tabular}{lcccc}
LIZ (sites)  & m$_{wire}$ ($\mu_B$) & m$_{sub}$ ($\mu_B$)
             & $\Delta E_{wire}$ (Ry) & $\Delta E_{sub}$ (Ry) \\ \hline
 13   &  2.248321 & 5.375824 & -0.011791 & -0.032602 \\
 19   &  2.244615 & 2.252704 & -0.010016 & -0.003562 \\
 43   &  2.232509 & 1.195110 & -0.008607 & +0.002313 \\
 55   &  2.223575 & 1.198589 & -0.007790 & -0.001625 \\
 79   &  2.223889 & 1.186796 & -0.007556 & -0.002480 \\
 87   &  2.222924 & 1.102650 & -0.007771 & -0.001743 \\
\end{tabular}
\end{ruledtabular}
\end{table}

\begin{figure}[floatfix]
\caption{Convergence with LIZ of the difference E$_{FM}$ - E$_{AFM}$ for
3d monowires on Pt ($311$) substrate.}
\vspace{7mm}
\begin{center}
\includegraphics[angle=-00,width=0.45\textwidth]{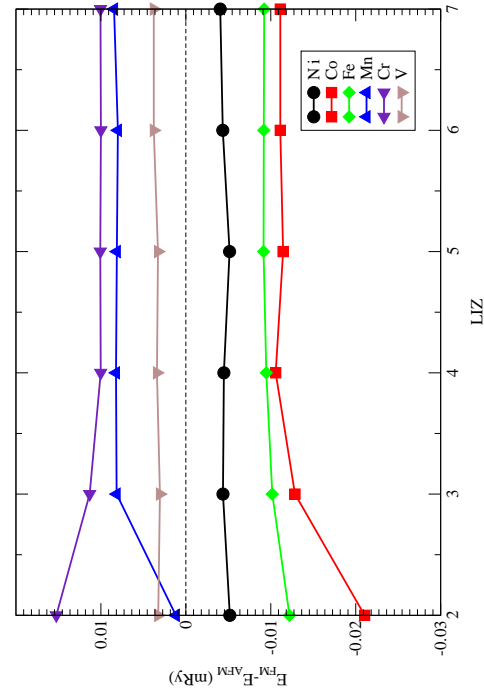}
\end{center}
\label{fig:convE}
\end{figure}

Finally we show in Fig. \ref{fig:convE} the convergence of the quantity
E$_{FM}$ - E$_{AFM}$ for all the 3d monowires from V to Ni deposited on
a Pt ($311$) substrate. Clearly, the convergence is with LIZ is sufficient
that one can state clearly which solution is preferable for each wire.

\section{Magnetic moments and magnetization energies}

We now turn to the trends of wire moment and energetics versus the d band filling. In Fig. \ref{fig:moments}
we show the wire, substrate, and vacuum moments for the elements V to Ni, Ti and Sc being found to display
no magnetic solutions for the monowire geometry considered here. As was found by Bl\"ugel \cite{Blugel92b} in 
the case of monolayers of 3d elements we find stable FM and AFM solutions for all monowires. This indicates
again that the monowire forms a good local moment system and particularly so for the Mn, Fe, and Co wires where
the moments of the FM and AFM configurations are almost identical.

\begin{figure}[floatfix]
\caption{Energy difference E$_{FM}$ - E$_{AFM}$ for 3d monowires on Pt ($311$) (open symbols) and ($533$)
(full symbols) vicinals.
Results shown are for $LIZ = 5$.}
\vspace{7mm}
\begin{center}
\includegraphics[angle=-00,width=0.45\textwidth]{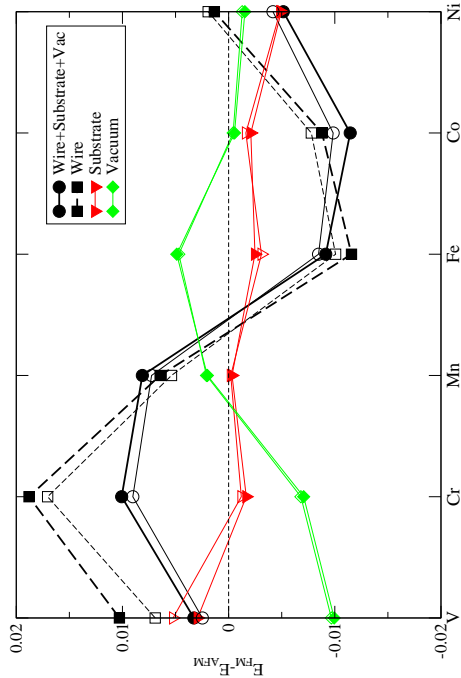}
\end{center}
\label{fig:energy}
\end{figure}

The total moment of the Pt substrate on the other hand is sensitive to the magnetic configuration
of the monowire, and changes from around 1 $\mu_B$ to almost zero when the wire configuration changes
from FM to AFM. One can note that the substrate moment associated with the FM monowire 
has a node between Cr and Mn, so that only for V and Cr does the substrate couple anti-ferromagnetically
to the FM monowire. The large polarization of the substrate associated with the FM configuration is
to be expected considering work on 3d impurities in Pd and Pt
where a large polarization cloud of several $\mu_B$ was observed experimentally and theoretically. This
being due to the fact that bulk Pt is near a magnetic instability. 
In Fig. \ref{fig:momRS} we show the total
magnetic moment of each principle layer of the ($533$) and ($311$) vicinals, the decay of the moment
is fairly slow and one can observe that nearly all layers of the Pt slab have become polarised.
On the other hand, the absence of a significant polarization of the substrate for the 
AFM monowire configuration indicates that the magnetic coupling is strongly ferromagnetic for 
the Pt vicinal surface.
The magnetic moment associated with the vacuum sites is seen to be largest earlier in the d series, which
is expected due to the larger spatial extent of the d orbitals for the early 3d transition metals. 

We note that the values found here for the Co wire spin moment and associated substrate moments are 
in close agreement with the very similar vicinal studied experimentally by Gambardella 
\cite{Gambardella02} {\it et. al}.

The quantity $\Delta E = E_{FM} - E_{AFM}$ is expected to possess two nodes across the d series and
in Fig. \ref{fig:energy} it can be seen that the second of these nodes occurs between Mn and
Fe, whereas the first would likely be seen between Ti and V if the Ti monowire was magnetic.
Thus we find that the V, Cr, and Mn monowires prefer and AFM structure whereas Fe, Co, and Ni
monowire prefer the FM structure. 
Turning now to the site projected energies one can see that the contribution to $\Delta E$ from the
vacuum and substrate sites is rather large away from the middle of the d-series. The origin of
the importance of this terms lies in the very different polarization of the surrounding depending
of the magnetic configuration of the wire. This might be expected to cause complications for
approaches based on extracting effective interactions for a Heisenberg model via the force theorem.
In this case one does perturbation theory on a particular magnetic structure, and thus would be
unable to take account of the change in energy coming from the change in polarization of the
environment.

\section{Conclusions}

We have studied monowire of 3d elements deposited on Pt ($311$) and ($533$) vicinal surfaces using a
real space order-$N$ method, and have investigated the convergence of various quantities as a 
function of the LIZ size. We demonstrate that the method is suitable for a quantitative
determination of the magnetization energies of the monowires, although the calculation of step-step
interactions would appear to be at the limit of what the method can achieve. Nevertheless, there
are many interesting magnetic structures, such as magnetic vicinals, which could be calculated with 
this approach.


\end{document}